\documentclass[12pt]{article}
\usepackage{amsfonts}
\usepackage{amssymb}
\usepackage{amsmath}
\usepackage{amsthm}
\usepackage{graphicx}

\begin{document}

\title{\bf Beyond dRGT as Mimetic Massive}

\author{\Large Alexey Golovnev${}^{a,b}$\\
${}^{a}${\it Asia Pacific Center for Theoretical Physics,}\\ 
{\it 67 Cheongam-ro, Nam-gu, Pohang 37673, Republic of Korea}\\
{\small alexey.golovnev@apctp.org}\\
${}^{b}${\it Faculty of Physics, St. Petersburg State University,}\\ 
{\it Ulyanovskaya ul., d. 1, Saint Petersburg 198504, Russia}\\
{\small agolovnev@yandex.ru}}
\date{}

\maketitle

\begin{abstract}

An interesting proposal has recently been made to extend massive gravity models beyond dRGT by a disformal transformation of the metric. In this Letter we want to note that it can be viewed as a mimetic extension of dRGT gravity which enormously simplifies the Hamiltonian  analysis. In particular, pure gravity sector is equivalent to the usual dRGT gravity coupled to a constrained scalar field. And we also give some comments about possible matter couplings.

\end{abstract}

\section{Introduction}

Gravitational interactions govern the evolution of our Universe, and the correct theoretical understanding of gravity is indispensable for a reliable description of cosmological evolution. General relativity is currently our best theory of gravity, and even at cosmological scales it allows to explain observational data in the framework of the standard cosmological model. However, a deeply disappointing feature is that some 95\% of the energy budget of the Universe must reside in the totally mysterious form of Dark Matter and Dark Energy.

It is not inconceivable though that our theory of gravitational interactions requires certain amendments in the infrared regime which would hopefully allow to ameliorate the cosmological puzzles. It gave rise to a whole new field of modified gravity theories which is very interesting even if only for complicated problems it makes the theorists to face. In particular, giving the graviton a mass generically results in a theory with a ghost \cite{BD}. It took some time to arrive at the model which avoids the ghost beyond the linear level in weak gravity regime.

Non-linear ghost-free massive gravity is relatively new and very active \cite{Claudia}. It turned out that ghost-free Lagrangians (dRGT) can be written down as
\begin{equation}
\label{act}
S=-\int \left(R(g)+\sum_i \beta_i e_i(\sqrt{g^{-1}\eta})\right)\sqrt{-g}d^4 x 
\end{equation}
where  $\beta_i$  are the parameters of the ghost-free potential, and $e_i$ are elementary symmetric polynomials of the eigenvalues of a square root of the matrix $g^{\alpha\mu}\eta_{\mu\beta}$ constructed from the fiducial Minkowski metric $\eta$ and the physical metric $g_{\mu\nu}$ with scalar curvature denoted in the action by $R(g)$. 

It is a very remarkable fact that such ghost-free model has been found \cite{dRG}. However, it turned out that it does not seem to allow for reasonable cosmological models \cite{nocosm}, even after some further extensions \cite{qdilco}. Some other extensions potentially could be more successfull, however they bring the ghost back \cite{Glenn, meTr}. It is possible to construct viable models in bimetric setup \cite{viable} which is arguably better justified than having a fixed fiducial metric, also from fundamental  perspectives. However, whether one can do better in a massive gravity regime is also a perfectly natural question to ask.

There is a common lore statement, partially supported by various nice arguments, that dRGT massive gravity is the only possible massive gravity theory. As usual with no-go statements, they are very useful in challenging attempts to go around.

There has appeared a paper \cite{NimaLuca} which aims at building new viable massive gravities via the disformal transformation of the metric:
\begin{equation}
\label{def}
g_{\mu\nu}=C(\phi,X)\cdot{\tilde g}_{\mu\nu}+D(\phi,X)\cdot(\partial_{\mu}\phi)(\partial_{\nu}\phi)
\end{equation}
where $\phi$ is a new scalar field with
\begin{equation}
\label{X}
X\equiv{\tilde g}^{\alpha\beta}(\partial_{\alpha}\phi)(\partial_{\beta}\phi).
\end{equation}
The recipe is that we use the action (\ref{act}) with the understanding that the metric $g$ is given by equation (\ref{def}) with $\tilde g$ and $\phi$ as independent variables. Matter is supposed to be coupled to $\tilde g$, however it is mostly pure gravity what is discussed \cite{NimaLuca}  at the moment. 

In this Letter we want to remind that such a construction as in \cite{NimaLuca} is known under the name of mimetic gravity.  In turn, mimetic gravity is known to either be totally equivalent to GR in cases when the transformation (\ref{def}) can be viewed as an invertible change of variables, or be equal to GR with a very simple constrained scalar field which makes the Hamiltonian analysis of such models very straightforward. Therefore, the message is two-fold. On one hand, it shows that models of Ref. \cite{NimaLuca} are ghost free. On the other hand, they don't go that much beyond the standard dRGT.

Coupling to matter is something to be separately investigated. Massive gravity is very unfriendly to messing up with matter couplings, at least in bimetric versions \cite{bimcop}. Apparently though, covariant couplings to a single metric should be on the safe side. However, we will argue that coupling to the untransformed metric ${\tilde g}_{\mu\nu}$ is not a good idea, even without the massive deformation of gravity.

\section{Brief reminder of mimetic gravity}

Mimetic gravity appeared in the Ref. \cite{mimetic} in the special case of $C(X)=X$ and $D=0$ which means that a simple relation
\begin{equation}
\label{definition}
g_{\mu\nu}={\tilde g}_{\mu\nu}{\tilde g}^{\alpha\beta}(\partial_{\alpha}\phi)(\partial_{\beta}\phi)
\end{equation}
has been substituted into the Einstein-Hilbert action. 

It is then easy to show \cite{mimetic} that the equations of motion are of the standard Einstein gravity coupled to
a pressureless ideal fluid with energy-momentum tensor of the usual form $\rho(x)\cdot(\partial^{\mu}\phi)(\partial^{\nu}\phi)$
where the scalar field is subject to the constraint:
\begin{equation}
\label{fixednorm}
g^{\alpha\beta}(\partial_{\alpha}\phi)(\partial_{\beta}\phi)=1,
\end{equation}
and the initial energy distribution $\rho$ can be chosen arbitrarily.
Obviously, equations became more general because the separation of the conformal mode depends on derivatives of an independent variable, and therefore it goes beyond a mute invertible change of variables \cite{me}.

Building  upon an obviously equivalent action
\begin{equation}
\label{equvact}
S=-\int\left(R(g)+\lambda^{\mu\nu}\left(g_{\mu\nu}-{\tilde g}_{\mu\nu}{\tilde g}^{\alpha\beta}(\partial_{\alpha}\phi)(\partial_{\beta}\phi)\right)\right)\sqrt{-g}d^4x,
\end{equation}
it was shown \cite{me} that the model can be written down (via integrating $\tilde g$ out) as a pure GR with a constrained scalar:
\begin{equation}
\label{final}
S=-\int\left(R(g)+\lambda\left(1-g^{\mu\nu}(\partial_{\mu}\phi)(\partial_{\nu}\phi)\right)\right)\sqrt{-g}d^4x.
\end{equation}
Note that it will not be changed by adding the massive terms for $g$. The form of the additional energy-momentum tensor $\lambda_{\mu\nu}=\lambda(\partial_{\mu}\phi)(\partial_{\nu}\phi)$ comes out directly from variation of $\tilde g$ in the action (\ref{equvact}).

Later this consideration has been generalised \cite{Nathalie} to arbitrary disformal transformations of the metric (\ref{def}) with the result that, depending on invertibility, this is either GR, or mimetic gravity. 

\subsection{Coupling matter to ${\tilde g}_{\mu\nu}$}

Above we have only discussed the gravitational sector.
Of course, as is always the case, the full model depends also on how the matter coupling is treated. In mimetic gravity, the transformation (\ref{def}) is usually \cite{mimetic, Nathalie} applied to the full action of the theory, that is including the matter part. On the contrary, in the Ref. \cite{NimaLuca} it was assumed that matter couples to the untransformed metric ${\tilde g}_{\mu\nu}$. Now we want to  show that it won't come out well, even in massless gravity. 

Indeed, let us consider the simplest model with an additional scalar field coupled to the metric ${\tilde g}_{\mu\nu}$:
$$S=-\int\left(\sqrt{-g}\left(R(g)+\lambda^{\mu\nu}\left(g_{\mu\nu}-{\tilde g}_{\mu\nu}{\tilde g}^{\alpha\beta}(\partial_{\alpha}\phi)(\partial_{\beta}\phi)\right)\right)-\sqrt{-\tilde g}{\tilde g}^{\mu\nu}(\partial_{\mu}\chi)(\partial_{\nu}\chi)\right)d^4x.$$
As it stands, it means that the field $\chi$ is coupled to a metric which is yet undetermined since the overall factor in ${\tilde g}$ is not given by a constraint imposed by $\lambda^{\mu\nu}$. 

It already seems pathological. However, let's try to further proceed with the model at hand. Variation with respect to ${\tilde g}^{\mu\nu}$ yields
\begin{multline*}
\sqrt{-g}\left(\lambda^{\alpha\beta}{\tilde g}_{\alpha\beta}(\partial_{\mu}\phi)(\partial_{\nu}\phi)-\lambda^{\rho\sigma}{\tilde g}_{\rho\mu}{\tilde g}_{\sigma\nu}{\tilde g}^{\alpha\beta}(\partial_{\alpha}\phi)(\partial_{\beta}\phi)\right)\\
+\sqrt{-\tilde g}\left((\partial_{\mu}\chi)(\partial_{\nu}\chi)-\frac12 {\tilde g}_{\mu\nu}{\tilde g}^{\alpha\beta}(\partial_{\alpha}\chi)(\partial_{\beta}\chi)\right)=0
\end{multline*}
After taking trace, the first part vanishes (because there was invariance under rescalings of $\tilde g$ in this part of the action) while the second part gives $(\partial\chi)^2=0$, a constraint for the physical matter which means that the energy momentum tensor must be traceless. It places an undesirable constraint on the matter dynamics  unless its energy momentum tensor was automatically traceless. However, in the latter case there is no difference between couplings to $g$ and to $\tilde g$.

For mimetic gravity with a general transformation (\ref{def}) the situation is similar. If it is invertible, then we have a mere change of variables, and the model at hand is equivalent to general relativity \cite{Nathalie}, or to dRGT massive gravity. Of course, if the matter is coupled to ${\tilde g}_{\mu\nu}$, then in terms of the physical metric $g_{\mu\nu}$ its action would contain interactions with a scalar field $\phi$ via inverse transformation ${\tilde g}={\tilde g}(g,\phi)$. This scalar enters universally for all fields and, as such, it can be naturally considered as modified gravity. However, we note that dRGT structure remains intact, and a new scalar sector is simply added on top of that. Stability of this construction requires further investigation, but it would anyway be very hard to intepret it as transcending the limitations of dRGT.

If the transformation is not invertible, then ${\tilde g}$ is not fully determined in terms of $g$, and there is an unphysical constraint on the $\tilde g$-coupled matter unless, for this particular field, there was no difference between $g$- and $\tilde g$-couplings. Therefore, in mimetic (non-invertible) cases the $\tilde g$-coupling prescription \cite{NimaLuca} is problematic.

\subsection{Disformal transformations}

The general case of mimetic gravity goes the same way without introducing new physical contents to the model. Let us briefly discuss it without dependence on $\phi$ (for simplicity)
\begin{equation}
\label{definition}
g_{\mu\nu}=C(X)\cdot{\tilde g}_{\mu\nu}+D(X)\cdot(\partial_{\mu}\phi)(\partial_{\nu}\phi)
\end{equation}
since the language of the Ref. \cite{Nathalie} is a bit different from ours \cite{me}.

The action can be written as
\begin{equation}
\label{action}
S=\int\left(R(g)+\lambda^{\mu\nu}\left(g_{\mu\nu}-C(X){\tilde g}_{\mu\nu}-D(X)\cdot(\partial_{\mu}\phi)(\partial_{\nu}\phi)\right)\right)\sqrt{-g}d^4x.
\end{equation}

Variation with respect to $\lambda$ imposes the condition (\ref{definition}). Variation with respect to $\phi$ gives
\begin{equation}
\label{phieq}
\bigtriangledown_{\mu}\left(\left(D\lambda^{\mu\nu}+\left(C^{\prime}{\tilde g}_{\alpha\beta}+D^{\prime}(\partial_{\alpha}\phi)(\partial_{\beta}\phi)\right)\lambda^{\alpha\beta}{\tilde g}^{\mu\nu}\right)\partial_{\nu}\phi\right)=0.
\end{equation}
Using relation (\ref{definition}),variation with respect to $g$ produces the Einstein equation $G^{\mu\nu}=\lambda^{\mu\nu}$ with a source.
Finally, we vary the action (\ref{action}) with respect to ${\tilde g}$ and get
\begin{equation}
\label{lambda}
\lambda^{\mu\nu}=\lambda^{\rho\sigma}\left(\frac{C^{\prime}}{C}{\tilde g}_{\rho\sigma}+\frac{D^{\prime}}{C}(\partial_{\rho}\phi)(\partial_{\sigma}\phi)\right){\tilde g}^{\mu\alpha}{\tilde g}^{\nu\beta}(\partial_{\alpha}\phi)(\partial_{\beta}\phi).
\end{equation}

We can define a scalar quantity
\begin{equation}
\label{lambdascalar}
\lambda\equiv\lambda^{\rho\sigma}\left(\frac{C^{\prime}}{C}{\tilde g}_{\rho\sigma}+\frac{D^{\prime}}{C}(\partial_{\rho}\phi)(\partial_{\sigma}\phi)\right)
\end{equation}
which gives by virtue of (\ref{lambda}):
\begin{equation}
\label{lambda2}
\lambda^{\mu\nu}=\lambda{\tilde g}^{\mu\alpha}{\tilde g}^{\nu\beta}(\partial_{\alpha}\phi)(\partial_{\beta}\phi).
\end{equation}
Lowering the indices by the physical metric, $\lambda_{\mu\nu}\equiv g_{\mu\alpha}g_{\nu\beta}\lambda^{\alpha\beta}$, and using (\ref{definition}) and (\ref{X}) we  get the simple form of effective energy-momentum tensor
\begin{equation}
\label{lambdalower}
\lambda_{\mu\nu}=\lambda (C+DX)^2 (\partial_{\mu}\phi)(\partial_{\nu}\phi)
\end{equation}
which corresponds to pressureless ideal fluid as in the classical case.

Note that the locus of $C+DX=0$ is singular because the physical metric becomes degenerate, $g_{\mu\nu}V^{\nu}=0$ for $V^{\nu}={\tilde g}^{\nu\alpha}\partial_{\alpha}\phi$, and therefore it should be avoided.
We can  define $\rho(x)\equiv\lambda (C+DX)$ which brings the effective fluid stress tensor (\ref{lambdalower}) to the form $\lambda_{\mu\nu}=\rho\cdot (C+DX) (\partial_{\mu}\phi)(\partial_{\nu}\phi)$, and using a simple relation $\lambda^{\mu\nu}\partial_{\nu}\phi=\lambda X \cdot{\tilde g}^{\mu\nu}\partial_{\nu}\phi$ it turns the equation of motion (\ref{phieq}) into
$\bigtriangledown_{\mu}\left(\rho(x)\cdot{\tilde g}^{\mu\nu}\partial_{\nu}\phi\right)=0$.

Finally, we multiply equation (\ref{lambda2}) by $$\frac{C^{\prime}}{C}{\tilde g}_{\rho\sigma}+\frac{D^{\prime}}{C}(\partial_{\rho}\phi)(\partial_{\sigma}\phi)$$ and get
\begin{equation}
\label{theproblem}
\lambda=\lambda\left(\frac{C^{\prime}}{C}X+\frac{D^{\prime}}{C}X^2\right).
\end{equation}
Therefore, either $\lambda_{\mu\nu}=0$ and the physical metric obeys the standard Einstein equations, so that we are back to GR with decoupled $\phi$, or if the relation
\begin{equation}
\label{condition}
C^{\prime}X+D^{\prime}X^2=C
\end{equation}
is satisfied (as was in the classical case), then $\lambda$ is arbitrary, and again we have an extra contribution of a pressureless ideal fluid. 

\section{Beyond dRGT gravity as massive mimetic model}

Now we apply the mimetic approach to the newly proposed \cite{NimaLuca} beyond dRGT gravity. Let us consider the simplest classical case of  $C(X)=X$ and $D=0$. One can repeat the calculations of the Ref. \cite{mimetic} for the action (\ref{act}) with the substitution (\ref{definition}). There are no changes exept that we have the dRGT massive term instead of unspecified matter action in the Ref. \cite{mimetic}. Therefore equations of motion would come out precisely the same with $G^{\mu\nu}-T^{\mu\nu}$ being substituted by $G^{\mu\nu}+\sum_i \beta_i Y^{\mu\nu}_i$ where $Y^{\mu\nu}_i$ is the variation of $\sqrt{-g} e_i(\sqrt{g^{-1}\eta})$ with respect to $g_{\mu\nu}$. It means that the model is nothing but the pressureless ideal fluid in dRGT massive gravity.

Another way, in spirit of the Ref. \cite{me}, is to consider the equivalent action
\begin{equation}
\label{actmim}
S=\int \left(R(g)+\sum_i \beta_i e_i(\sqrt{g^{-1}\eta})+\lambda\left(1+g^{\mu\nu}(\partial_{\mu}\phi)(\partial_{\nu}\phi)\right)\right)\sqrt{-g}d^4 x.
\end{equation}
Obviously, the constraint which removes the Boulware-Deser ghost is present. Indeed, we remind that the classical proof \cite{HR} is based upon a (linear in lapse $N$) redefinition of shifts $N^i$ after which the action is linear in the lapse which therefore serves as a source of constraint for the spatial metric $\gamma_{ij}$. 

In the current case (\ref{actmim}) we see that the canonical momentum of the scalar field in the usual ADM variables \cite{HR} is 
$$\pi_{\phi}=\frac{2\lambda \sqrt{\gamma}}{N}\left({\dot\phi}+N^i\partial_i\phi\right),$$
and the contribution to the Hamiltonian density reads
$${\mathcal H}=\sqrt{\gamma}\left(\frac{N\pi_{\phi}^2}{4\lambda\gamma}-\frac{\pi_{\phi}N^i\partial_i\phi}{\sqrt{\gamma}}+\lambda N\gamma^{ij}(\partial_i\phi)(\partial_j\phi)\right)$$
which does not compromise linearity in the lapse, even after a linear in lapse redefinition of shifts is performed.

With a general disformal transformation, the gravitational sector is either equivalent to the initial gravitational model, be it massless or massive, or it adds a pressureless ideal fluid much the same way \cite{Nathalie} as in the classical case of relation (\ref{definition}). Of course, the physical conclusions should not be changed. However, elaborating it out explicitly requires some work. 

Indeed, the action (\ref{final}) can be obtained from the action (\ref{equvact}) by substituting an obvious ansatz ${\tilde g}_{\mu\nu}=\alpha(x)\cdot g_{\mu\nu}$ and denoting $\lambda\equiv\lambda^{\mu\nu}g_{\mu\nu}$. Dependence on $\alpha(x)$ automatically disappears reflecting the non-invertibility of the transformation  (\ref{definition}). In the general case one has to use a more complicated ansatz of ${\tilde g}_{\mu\nu}=\alpha(x)\cdot g_{\mu\nu}+\gamma(x)\cdot(\partial_{\mu}\phi)(\partial_{\nu}\phi)$ with two remaining scalar Lagrange multipliers, $\lambda(x)\equiv\lambda^{\mu\nu}g_{\mu\nu}$ and $\mu(x)\equiv\lambda^{\mu\nu}(\partial_{\mu}\phi)(\partial_{\nu}\phi)$. Under certain circumstances, coefficients $\alpha$ and $\beta$ would not be fully determined by equations of motion with the gravitational model getting coupled to a constrained k-essence, akin to the field $\phi$ in the action (\ref{final}). We leave details for a future work.

\section{Conclusions}

We have seen that the model presented in the Ref. \cite{NimaLuca} is actually a mimetic extension of massive gravity if the transformation (\ref{def}) is non-invertible, or otherwise it reduces to the unaffected dRGT. 

This fact allows one to extend the Hamiltonian analysis of the Ref. \cite{HR} very easily. In particular, it is very simple to find the Hamiltonian constraint in the classical case of mimetic transformation (\ref{definition}). General disformal transformations are known to produce equivalent models, however an explicit Hamiltonian analysis directly in their terms requires more work to be done.

We also considered the new idea of the Ref. \cite{NimaLuca} about the matter coupling, namely to use the untransformed metric for that. It turns out that in invertible cases it adds a new scalar field into the action of matter, rather intricately but independently of the (unaffected) gravitational part. In non-invertible (mimetic) cases this type of coupling is problematic because it imposes undesired constraints on the matter fields.

We summarise by noticing that the models presented in the Ref. \cite{NimaLuca} are ghost-free. However, they essentially remain in the class of dRGT massive gravity. The question of whether we can go substantially beyond this class is still open. At least, disformal transformations do not offer much support to the claim of the Ref. \cite{NimaLuca} that "dRGT massive gravity is not unique."

\end{document}